\documentclass[sigconf]{acmart}

\newcommand{\nicki}[1]{\textcolor{black}{#1}}
\newcommand{\update}[1]{\textbf{}\textcolor{black}{#1}}

\newenvironment{myquote}%
  {\list{}{\leftmargin=0.15in\rightmargin=0.15in}\item[]}%
  {\endlist}
  
\AtBeginDocument{%
  \providecommand\BibTeX{{%
    \normalfont B\kern-0.5em{\scshape i\kern-0.25em b}\kern-0.8em\TeX}}}

\copyrightyear{2024}
\acmYear{2024}
\setcopyright{acmlicensed}\acmConference[CHI '24]{Proceedings of the CHI Conference on Human Factors in Computing Systems}{May 11--16, 2024}{Honolulu, HI, USA}
\acmBooktitle{Proceedings of the CHI Conference on Human Factors in Computing Systems (CHI '24), May 11--16, 2024, Honolulu, HI, USA}
\acmDOI{10.1145/3613904.3642617}
\acmISBN{979-8-4007-0330-0/24/05}

\begin{document}

\title[Saharaline]{Saharaline: A Collective Social Support Intervention for Teachers in Low-Income Indian Schools}

\renewcommand{\shortauthors}{}

\begin{abstract}

\update{
This paper presents Saharaline, an intervention designed to provide collective social support for teachers in low-income schools. Implemented as a WhatsApp-based helpline, Saharaline enables teachers to reach out for personalized, long-term assistance with a wide range of problems and stressors, including pedagogical, emotional, and technological challenges. Depending on the support needed, teachers' requests are routed to appropriate domain experts--- staff employed by educational non-profit organizations who understand teachers' on-the-ground realities---who offer localized and contextualized assistance. Via a three-month exploratory deployment with 28 teachers in India, we show how Saharaline's design enabled a collective of diverse education experts to craft and deliver localized solutions that teachers could incorporate into their practice. We conclude by reflecting on the efficacy of our intervention in low-resource work contexts and provide recommendations to enhance collective social support interventions similar to Saharaline.}

\end{abstract}

\author{Rama Adithya Varanasi}
\affiliation{%
  \institution{Department of Information Science, Cornell University}
  \city{Ithaca}
  \state{NYC}
  \country{USA}
}
\author[N. Dell]{Nicola Dell}
\affiliation{
 \institution{Department of Information Science, the Jacobs Institute, Cornell Tech}
 \city{New York}
 \state{NY}
 \country{USA}
}
\author[Aditya V.]{Aditya Vashistha}
\affiliation{
 \institution{Department of Information Science, Cornell University}
 \city{New York}
 \state{NY}
 \country{USA}
}



\maketitle

\section{Introduction}

Teachers in low-income schools\footnote{We use the term "low-income schools" to encompass schools that serve low-income communities, including government schools and affordable private schools that operate with limited resources.}, particularly in the Global South, contend with resource-constrained environments~\cite{Susan2005who, Mulkeen_2010}. They teach large classes with immense learning variability among students \cite{Alcott_Rose_2015, Nath_2012}, prepare teaching materials and pedagogical strategies using scarce infrastructure \cite{Marishane_2014, wbpubs:30920}, and have limited opportunities for upskilling \cite{Govmda_Josephine_2005, Ogunniyi_Rollnick_2015}. Together, these burdens contribute to teachers’ occupational stress, with increasing evidence pointing to teachers’ lack of job satisfaction \cite{Basak_Ghosh_2011}, reduced motivation \cite{bold2017}, and burnout \cite{Shukla_Trivedi_2008, Varanasi_2023}.

Increasingly, educational technology organizations and researchers expect teachers to overcome resource limitations by reconfiguring their personal smartphones for work, creating additional stressors. For example, research in low-income private schools in India showed that teachers are expected to use smartphones to find new ways to prepare, teach, and engage students \cite{gesu2021, varanasi2019}. At the same time, school management uses the same smartphones to increase teacher surveillance and stretch teachers' work responsibilities beyond their work hours \cite{varanasi2021Investigatinga}. These conflicting practices around smartphones have further contributed to new forms of stress, including technostress (stress induced by technology) and burnout. Moreover, with the COVID-19 pandemic, technologies that were previously considered optional have became critical infrastructure for curriculum preparation and delivery, and other administrative tasks \cite{ravi2021}. The subsequent top-down shifts to hybrid work practices during and after the pandemic are further exacerbating the already demanding nature of the teaching profession, impacting teachers' overall occupational well-being and productivity \cite{Zhao_Wang_Wu_Dong_2022}. 

In such challenging circumstances, social support provides a crucial avenue for teachers to cope with stress and improve their occupational well-being \cite{Kyriacou_2001}. Social support is a specific coping mechanism where an individual leverages their social ties and community resources to receive assistance \cite{Kim_Oxley_Asbury_2022}. \update{In the context of teaching, social support plays a vital role in enabling teachers to manage their occupational well-being, particularly given the constant emotional labor demanded by their jobs \cite{kinman2011emotional}. However, providing holistic and personalized social support in low-income schools poses several challenges. 
For example, establishing new social support structures may require school leadership to re-allocate resources away from other programs and initiatives, or demand resources and infrastructure (e.g., time, personnel) that schools simply do not have. 
Consequently, on-the-ground efforts in low-income schools are often constrained to general professional development initiatives that lack personalization \cite{gavade2023}. 
As a result, teachers must often}
rely on self-developed support structures that add additional pressure to their already overburdened responsibilities. For example, peer-based social support initiated by teachers, while effective, tends to increase their peers' workload \cite{gavade2023}. Moreover, these structures also skew towards types of support that are easier to seek (e.g., pedagogical) than those that are more challenging (e.g., emotional) \cite{gavade2023}. \update{There is thus a need for social support structures that provide holistic, personalized assistance to teachers in ways that do not strain the current ecosystem}. 

To address this need, we created Saharaline\footnote{In Hindi, `Sahara' translates to `support.'}, a hybrid\footnote{We use the term `hybrid' to denote that the intervention facilitates both online and in-person interaction.} sociotechnical intervention to improve teachers' occupational well-being in low-income schools through \textit{collective social support}. We define collective social support as a form of support that unites different individuals in the communities where workers are embedded to provide comprehensive, localized, and longitudinal assistance. Using this philosophy, Saharaline was deployed on WhatsApp to bring different educational non-profit organizations (referred to as support organizations, hereafter) that work in school environments together to provide social support to teachers (see Figure \ref{architecture}). Teachers across six different schools in India were informed about the Saharaline service, and they accessed it by sending messages or calling a WhatsApp number. A \textit{facilitator} managing Saharaline's WhatsApp interface captured their information. Based on the data, a \textit{caseworker} working in a support organization within the community was mapped to the teacher, who recorded the teacher's key problems and shared them back with facilitators. Depending on the nature of the problem, facilitators then assigned a remote expert with relevant expertise (occupational mental health, technology education, or pedagogical and content knowledge) to develop solutions to the teacher's problems. Subsequently, the facilitator conveyed these solutions to caseworkers who communicated them to teachers. Saharaline was designed to provide multiple rounds of such support to teachers, \update{in ways that were manageable within teachers' and experts' current workloads.}


We conducted a three-month pilot deployment of Saharaline in India, focusing on the experiences of 28 teachers and 11 support organization personnel (facilitators, caseworkers, and experts) as they received and provided various forms of social support, respectively. Our findings indicate that Saharaline attracted diverse teachers with many problems, including those with emotion-focused issues who were otherwise hesitant to seek support through their usual channels. By building rapport with teachers, caseworkers successfully surfaced key problems along with their local practices and communicated them to Saharaline experts. The remote experts collaborated with caseworkers to craft effective and contextual solutions while carefully navigating the complexities of teachers' work practices. 

We critically reflect on our findings to discuss Saharaline's distributed, versatile approach in delivering collective social support. Our discussion aims to understand the benefits and challenges it offers to low-resource working communities, such as teachers. We also unpack the decentralized knowledge production practices we observed and provide design recommendations on integrating emerging technologies to increase the efficiency of the intervention. Finally, we conclude by reflecting on the complexities of providing social support via WhatsApp, a piece of technology that can itself be a source of stress for teachers. In sum, our paper makes four contributions: 
\begin{itemize}
    \item We show how collective social support was effective in surfacing a wide range of teacher stressors.
    \item We show how most teachers (n=25) who reached out to Saharaline did so for multiple rounds of social support, successfully engaging in longitudinal support during the three month deployment.
    \item We present different support pathways created by Saharaline stakeholders to develop scalable social support structures across six schools in three Indian states.
    \item We discuss key takeaways to inform future collective social support interventions and increase their positive impact in low-income working communities.
\end{itemize}
\section{Background and Related Work}

\subsection{Understanding Dimensions of Social Support}

\citet{shumaker1984Theory} define social support as a process of ``exchanging resources'' between individuals where either the provider or receiver explicitly intends to improve the well-being of the receiver. It serves to aid individuals facing stress and negative well-being \cite{lazarus20Stress, Jolly_Kong_Kim_2021}. While these studies focused on stressors in one's personal life, research has shown similar findings for occupational well-being \cite{ganster1986role,Liu_Aungsuroch_2019}. In work settings, social support takes various forms, including peer support \cite{Mercieca_Kelly_2018}, vocational interventions \cite{McGrath_2020}, assistive programs \cite{Kirk_Brown_2003}, mentorship and coaching \cite{Jones_Woods_Guillaume_2016}. 

Regardless of their manifestations, social support structures are typically governed by three underlying dimensions ~\cite{wills1991social,taylor2011social}. The first is the \textit{function} of support in an individual's life \cite{house1981Work}. Support can be \textit{problem-focused}, providing assistance to ``manage or alter the problem causing the stress'' \cite{lazarus1984stress}. Alternatively, support can be \textit{emotion-focused}, intended to help regulate emotional responses to challenges \cite{lazarus1984stress}. Examples include providing emotional warmth and care \cite{lazarus20Stress}, network support \cite{Cutrona_Suhr_1992}, and boosting esteem \cite{Cutrona_Suhr_1992}. The second dimension involves an individual's \textit{perception} and evaluation of their workplace social support \cite{sarason1991perceived}. Research has consistently shown a strong positive link between perceived social support and occupational well-being \cite{Park2004, turner1981social}. 
The third dimension pertains to the \textit{structure} of support, determined by how individuals embed themselves and engage with social networks \cite{Chronister_2006, Jolly_Kong_Kim_2021}. This support can take two main forms: \textit{individual} support, where a person receives one-on-one assistance; a common example being therapy or mentorship \cite{Vincent2020}. Alternatively, the support can be \textit{collective}, where an individual receives support as a result of cooperation or collaboration from multiple people \update{\cite{Archer_2012}}. What distinguishes collective support from individual support is the combined effort of several providers to deliver social support \update{\cite{Bayraktar_2019}}. In the following sections, we explore individual and collective social support around work and technology, especially within teaching context to make a case for the effectiveness of collective social support for low-resource teachers to enhance their occupational well-being.

\subsection{Individual Social Support: Technology, Work \& Teaching}
The widespread availability of Internet connectivity has opened up avenues for technology-mediated individual social support, via channels such as emails \cite{Hutson_Cowie_2007}, forums \cite{Turner_Grube_Meyers_2001}, blogs \cite{rains2011}, and online conversations, including remote therapy \cite{Boos_2022}. Despite early research showing mixed results \cite{kraut1998Social, shaw2002defense}, the proliferation of devices has expanded opportunities for informational \cite{Coursaris_Liu_2009}, emotional \cite{Buis_2008}, and network \cite{Ashley2011} support provided via online personal networks \cite{han2016}.

HCI research has also studied individual social support practices within specific work domains, including upskilling \cite{Dill2016} and career transitions \cite{burke2013, radhika2020, Dillahunt21}. This knowledge has equipped HCI researchers to examine social support in demanding professions, particularly emotionally taxing work such as teaching. Teachers have to constantly regulate their emotions while working in isolation and contending with frequent stress \cite{Madigan_Kim_2021}. The integration of technology into education has further introduced new forms of technology-enabled stress, putting teachers at the risk of burnout \cite{Zhao_Wang_Wu_Dong_2022}. For example, during COVID-19, transitions to hybrid teaching exacerbated teachers' isolation and stress \cite{Gonzalez_2021, Westphal_2022}. 

In the Global South, where our research is situated, these challenges are amplified for several reasons. Teachers rely heavily on smartphones for teaching, given their affordability and adaptability, as opposed to laptops or desktop computers \cite{cannanure2020}. Moreover, these smartphones are often teachers' personal devices used for work, introducing complexities not encountered by educators using primarily school-provided technological infrastructure \cite{varanasi2020}. Lastly, the COVID-19 pandemic also forced teachers, many of whom had limited experience with remote teaching, to embrace hybrid teaching approaches using smartphones \cite{ravi2022, Singh_2023}. Early studies suggest that this transition has led to reduced confidence, increased burnout, and higher attrition rates \cite{varanasi2021Investigatinga}. Robust social support systems could be a critical resource for teachers to combat isolation, enable professional development, and enhance occupational well-being \cite{Richard2011}. Research in this area has explored the effectiveness of peer interactions among teachers in facilitating informational, instrumental, and emotional support through online and offline mediums, resulting in reduced stress levels among educators \cite{dewert2003safe, kinman2011emotional, paulus2008can}. 

However, teachers' access to peer and management-established social support is contingent on the availability of school resources \cite{Westphal_2022}. In resource-constrained schools, the absence of resources puts the onus on teachers to establish and maintain their own support networks. For instance, \citet{gavade2023} showed how, in low-income Indian schools, teachers' formal channels for seeking support from school management were virtually non-existent, forcing teachers to use their own means (e.g., smartphones) to seek support. Additionally, formal support, requires significant resources from providers (e.g., management), increasing burdens and hampering functionality of support programs \cite{gavade2023}. 

\subsection{Collective Social Support in Work Settings}
In contrast to individual social support, \update{collective social support distributes responsibilities across different roles within the ecosystem, encouraging multiple individuals to collaborate, coordinate, and deliver support. Collective social support can manifest as \textit{internal}, wherein various roles, such as peers, management, and other stakeholders come together to assist individuals. This form of support} has become common in roles involving emotional labor, such as hospitality workers \cite{Wu_Chen_2019}, healthcare workers \cite{Tixier2009}, social workers \cite{Matthie2016}, gig workers \cite{Uttarapong_Bonifacio_2022} and teachers \cite{varanasi2020}. With social media communication (e.g., Facebook and WhatsApp) becoming prominent in workspaces, virtual interactions (e.g., likes, reactions, and emojis) have increased social connectedness \cite{varanasi2018, Carr_Wohn_Hayes_2016} and have provided new pathways for workers to reach multiple people and seek support. While this form of support reduces the burden on management by involving different roles within the working community, the overall strain is still significant for a resource-constrained community \cite{Oreglia_2020}.    

\nicki{To overcome this issue, collective social support efforts can also leverage individuals {\it external} to the community \nicki{\cite{Bayraktar_2019, Archer_2012}}}. One strand of research that shapes this notion of collective support is the \textit{collaborative care} (sometimes called collective or cooperative care) movement in healthcare \nicki{\cite{Archer_2012}}. Collaborative care mobilizes medical \nicki{experts and caregivers across different domains to collaborate and provide social support tailored to an individual's needs. Such collaborative care leverages coordination among multiple specialists, thereby reducing strain not only on specific medical roles but also on the overall healthcare system}. This type of support has been shown to improve individuals' well-being when compared to traditional support \cite{Archer_2012}. \update{However, involving different stakeholders brings about its own challenges. For instance, defining and maintaining clear boundaries for distinct roles can be challenging \cite{Vanderlip2016}. Diversifying care also introduces unequal burdens in maintaining standardized care pathways, posing challenges to effective social support \cite{Moise_2018}.
} HCI research has built on these results to understand how sociotechnical solutions might include non-specialists like family members in providing collaborative care for vulnerable populations \cite{Soubutts2021, suh2020}. \update{For instance, \citet{amy2020} found that the learnability and ease of use of technological tools are critical elements for the success of providing and receiving collective support. In the long run, such instances of successful support translate into improved coping mechanisms, directly impacting individuals' abilities to regulate their stress \cite{caldeira2017}.} 

\update{Another strand of research has examined \textit{collective social support} in resource-constrained settings \cite{Arueyinghoal_2023}}. Similar to collaborative care, this form of social support engages actors outside the community to provide support to reduce stress on the system  \update{\cite{ismail2022}}. \nicki{For example, in Côte d’Ivoire, families and community members volunteered to complement teachers' limited resources by providing informal education support for their children \cite{Madaio20}}. This involved parents, siblings, relatives, and neighbors sharing the responsibilities of informal teaching. \update{Involving community members who were invested in the children's future allowed cost-effective implementation of support activities \cite{Madaio20}.} However, a key issue with both these strands of research is their narrow focus on the individuals or communities where workers (medical specialists, teachers) provide support (to patients, students), potentially overlooking the well-being and support needs of the workers themselves  \update{\cite{varanasi2021Investigatinga}}. These initiatives also focused on specific types of issues (mental health, learning outcomes) and do not cover other stressors that might be equally important for occupational well-being \update{\cite{Singh_2023}}. 

Limited research in the Global South, including health \cite{Ismail_Kumar_2019} and education \cite{varanasi2020}, has focused on working communities, particularly the role of intermediaries in providing collective support to workers. \update{To address the limited resources and substantial variability in the digital capabilities of these communities, interventions in this context have adopted an "asymmetrical" design \cite{rahman2021}. In this design, user-facing technologies are developed to be simple and straightforward, with more complex designs and heavier load on the back-end to scale the support \cite{rahman2021}. For example, \citet{Cannanure_2020} implemented a conversational agent in the backend that teacher trainers could leverage to consolidate common professional development topics in which  teachers were interested.} However, at a high-level such research has predominantly aimed at addressing specific work-related challenges (e.g., productivity through professional development) rather than fostering comprehensive occupational well-being \update{\cite{Oreglia_2020}}. Nevertheless, this literature underscores the need for collective social support practices for low-resource working communities, especially in the Global South.

Our study expands this literature by designing and deploying Saharaline, a hybrid collective social support system that is explicitly tailored towards \textit{workers} (i.e. teachers) as opposed to the community they serve (i.e. students). In this novel context, multiple support organizations specializing in different forms of social support, such as work-related specialization (pedagogy, content), technology, and mental health, took on distinct roles to provide social support that was \textit{personalized} and \textit{holistic}, focusing beyond work-related challenges and covering major aspects of \textit{occupational well-being}. Through a three-month deployment of Saharaline in low-resource working communities of teachers, our research explores the feasibility of providing longitudinal collective social support via a commonly available technology platform like WhatsApp. In doing so, we sought to answer two research questions: 

\noindent\textbf{RQ1}: \textit{What are the experiences of support organization personnel in providing collective social support through a hybrid sociotechnical intervention?} \\
\textbf{RQ2}: \textit{How do teachers use a hybrid collective support intervention to improve their occupational well-being?}

\section{Methods}
Here, we describe our IRB-approved research methods, including Saharaline's design, operation, and details of its three-month deployment in low-income Indian schools (March to May 2022).  

\subsection{The Educational Support Organization Ecosystem in India}
Educational support organizations are non-governmental organizations or social enterprises that collaborate with low-income private and government schools to enhance educational outcomes through capacity-building initiatives. \nicki{In India, where educational non-profit organizations hold the largest proportion among all non-profit entities \cite{Niti}, such collaborations are common. In these collaborations, the support organizations can engage at three levels }(a) At the student level, they focus on interacting with students to improve learning outcomes (Support Organization 3 in our study, Table \ref{tab:sup-org} in the appendix); (b) At the teacher level, they train teachers through workshops to enhance their capacities (Support Organization 1 in Table \ref{tab:sup-org}); (c) At the management level, they collaborate with school management to shape school-level policies and strategies (Support Organization 2 in Table \ref{tab:sup-org}). 

\update{Regardless of the level at which they operate, support organizations typically employ two types of personnel. The first works off-site at the organization's headquarters, including subject matter experts, leadership, and the management team, situated outside the school community. These people typically possess advanced degrees (e.g., education psychology) and contribute their global awareness and domain-specific knowledge to support organizational efforts. 
The second type of personnel works on-site in schools. These people are recruited from the same local community where the schools are located. They generally possess a bachelor's degree and bring a wealth of local knowledge and experience gained through their work in the community. In this capacity, they serve as intermediaries between the personnel at the headquarters and the stakeholders on the ground. Thus, organizations comprise members from both within and outside the school community, enabling the integration of global expertise with local context. Support organizations usually visit school clusters or government offices to advertise their services. If there is interest, the organization and school sign a Memorandum of Understanding lasting anywhere between 2 to 5 years.}

\subsection{Design Approach and Objectives}
Saharaline was developed collaboratively by three researchers who study teacher work practices, three support organization members, and four experienced teachers from low-resource schools. Beyond sharing a common motivation to positively impact low-income schools, the three support organizations also shared an overlapping objective of understanding and addressing challenges teachers face. All organizations had previously collaborated with the researchers, actively engaging in exploratory projects that focused on teachers' sociotechnical practices. Over a four-week period, all the stakeholders came together to collectively design the overall architecture of Saharaline using a scenario-based design approach  \cite{Carroll_1997,schuler1993participatory} (see Table \ref{tab:design-activity} in the appendix). The first week involved reviewing literature, including relevant projects in HCI (e.g., \cite{varanasi2019, cannanure2020}), teacher education (e.g., \cite{Govmda_Josephine_2005, Ogunniyi_Rollnick_2015}), and decolonial research (e.g., \cite{pendse2022, smith2021decolonizing}), and sharing stakeholders' lived experiences on the ground; this utilized a seminar format that resulted in story scenarios. \nicki{Example story scenarios included teachers' experiences of using smartphones for preparation, teachers' conversations with parents of children with learning disabilities, and teachers' interactions with their peers in the staff room.}

In the second and third week, we conducted co-creation workshops with all the stakeholders to translate the story scenarios into problem scenarios. \nicki{For instance, stakeholders used the finalized stories to derive specific problem scenarios where smartphone use impacted their preparation process. The problem scenarios that captured their pain points were then used to} finalize the key design principles. During these workshops, we used affinity diagramming and expert walkthroughs to capture diverse stakeholder perspectives. The final week involved translating the finalized design principles into a high-level design architecture and obtaining feedback on specific implementation pathways in an attempt to anticipate and address potential challenges that might arise. All activities were moderated by a researcher and a support organization personnel. We now briefly discuss the four resultant design objectives for Saharaline that materialized from this collaboration, followed by the overall architecture and implementation.

The first design objective was to anchor Saharaline's support services in teachers' lived experiences \cite{Mignolo_Walsh_2018}, allowing Saharaline to focus on teachers' challenges localized to their communities. To achieve this objective, we sought collaboration of support organizations (or educational NGOs) as social support providers. These organizations work with low-resource schools to build capacity at various levels (e.g., students, teachers, and management) through workshops, seminars, and professional development classes. While none specifically targeted teachers' \update{overall occupational well-being in their capacity development efforts}, they were aware of the specific sociotechnical and sociocultural challenges teachers faced in schools. \update{Recognizing the cascading impact of addressing teachers' issues on their own capacity development initiatives, these organizations naturally expressed interest in engaging with individual teacher problems through Saharaline}. Their awareness of teachers' local experiences enabled them to better understand teachers' stressors, which traditional support systems did not adequately address.

The second objective was to decentralize Saharaline's support providers and deliver collective social support by fostering collaboration among multiple support organizations \cite{Ogunniyi_Rollnick_2015}. 
\update{Through this design objective, we wanted to identify support providers that brought their unique awareness of school community challenges and complementary expertise in teacher practices, as well as their pedagogical and content knowledge. Such diverse perspectives can reduce burden on a single organization by facilitating the co-creation of efficient support solutions between teachers and members of support organizations.} For example, a low-income school can have two support organizations working simultaneously, one working to empower the management to improve the school leadership and administrative capacities, and the other working with teachers to improve their pedagogical capacities. These organizations' functions overlap, leading to cooperation and assistance that can be leveraged to support teachers. At the same time, these organizations have different priorities, allowing them to bring their unique resources and expertise, while reducing their individual efforts. 

\update{The third objective was to make Saharaline's support accessible by offsetting the power differential~\cite{Lazem_2022} that traditionally exists between the management and teachers, which limits teachers' support seeking practices due to potential repercussions. As support organizations occupied a neutral position outside the conventional hierarchies of the school system, they did not carry the same complex power dynamics as school management. This allowed them to gain trust and engage in sensitive conversations, facilitating an in-depth understanding of teachers' challenges.}

\update{The fourth objective was to ensure that the work involved in using and running Saharaline would be manageable given stakeholders' current workloads. Specifically
(i) that teachers could fit it into their overburdened work schedules and (ii) that the intervention was manageable for support providers to deliver alongside their organization work. To achieve this, we developed Saharaline as a hybrid intervention, consisting of both online (i.e., WhatsApp-based helpline) and in-person interactions (i.e., meetings with teachers). }

We chose WhatsApp because it is a widespread and familiar tool in low-income schools due to its ease of use and high adoption rate among teachers in India.
The business version of WhatsApp is geared towards organizations, non-profits, and businesses, providing automation features (e.g., scheduled greeting messages, automated replies, predefined shorthand commands) and multi-user management of a single account. This set-up allowed us to scale the initiative across multiple schools in different communities. We complemented this technological choice with in-person support practices that could accommodate teachers' busy schedules and provide long-term support.

\begin{figure*}
 \centering
 \includegraphics[width=0.7\linewidth]{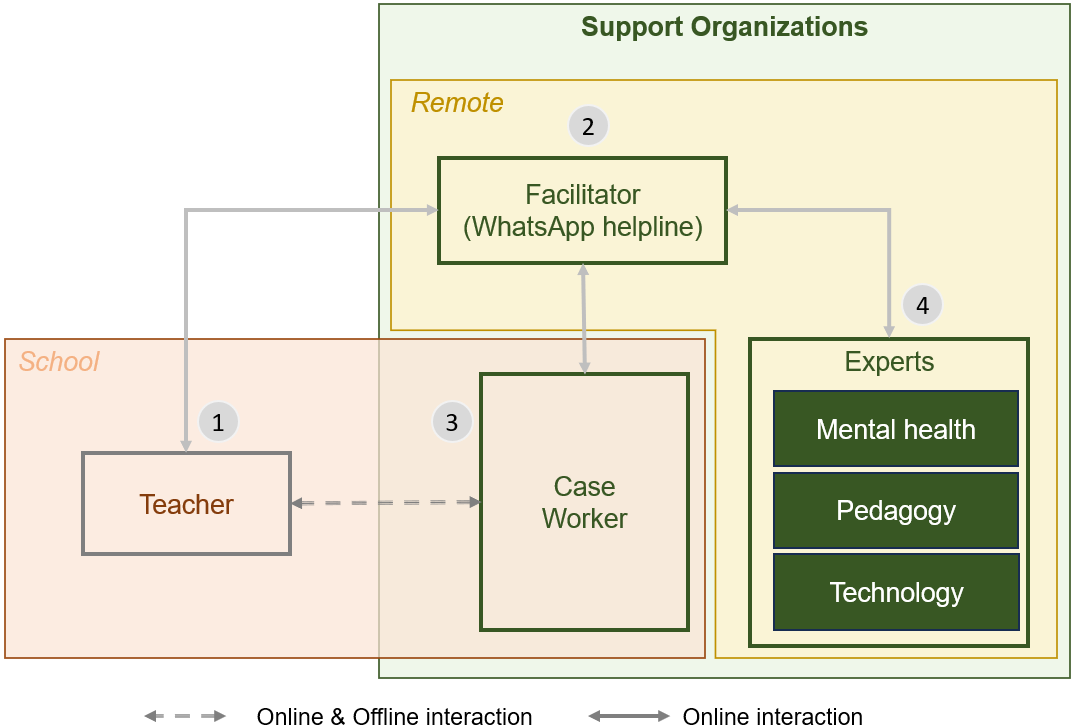}
  \caption{Saharaline architecture showcasing key stakeholders: (1) teachers in low-resource schools who contacted Saharaline, (2) facilitators who oversaw coordination and managed the helpline;  (3) caseworkers who interacted with teacher on behalf of Saharaline; (4) experts who provided the solutions. }
  \Description{Saharaline architecture showcasing key stakeholders, namely teachers in low-resource schools who contacted Saharaline facilitators who oversaw coordination and managed the helpline, caseworkers who interacted with teacher on behalf of Saharaline and experts who provided the solutions. }
  \label{architecture}
\end{figure*}

\begin{figure*}
 \centering
 \includegraphics[width=\linewidth]{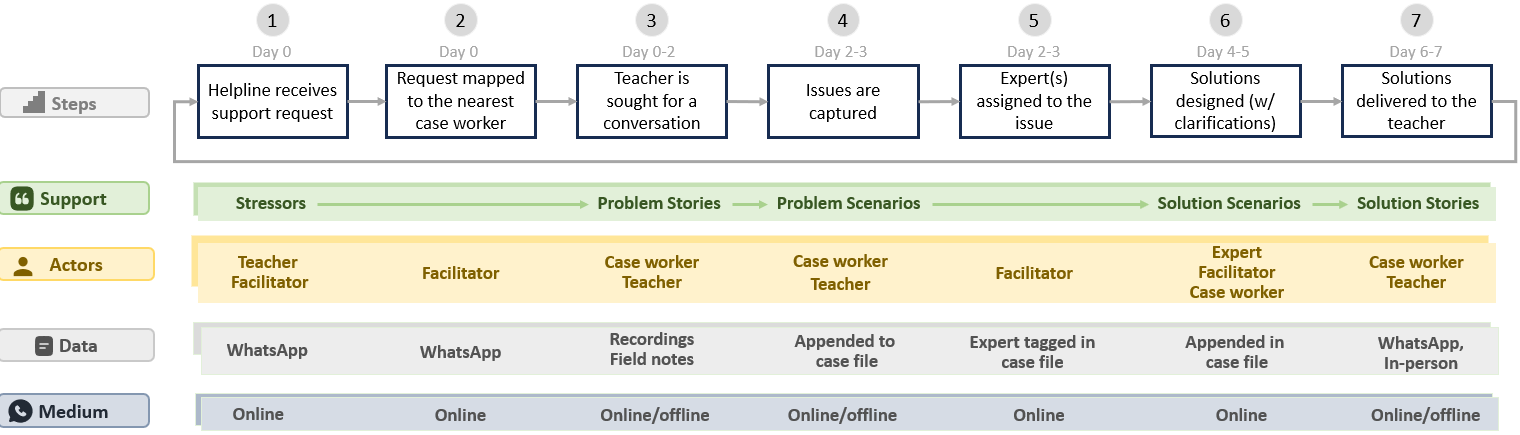}
  \caption{Breakdown of pathways providing collective social support through Saharaline; Support represents the conversion of stressor to solutions for each stage; Actors represent different stakeholders involved in each step; Data and medium showcase different mediums in which data is produced through interaction or knowledge production.}
\Description{Overview of Saharaline pathways being shown in five rows for seven stages of support. First row depicts support actions in each stage. Second row depicts actors involved. Third row depicts the data that is created and the fourth row shows the medium of support.}
  \label{workflow}
\end{figure*}

\subsection{Saharaline Operation}
\label{operation}

Figure \ref{architecture} shows Saharaline's architecture and stakeholders, while Figure \ref{workflow} details the steps involved in providing social support to teachers.  
To avail support, teachers contacted Saharaline through a WhatsApp message or called seven days a week between 8am and 8pm. Ensuring daily operation gave teachers the flexibility to engage with Saharaline at their convenience. Upon connecting to the helpline (step 1, Figure \ref{workflow}), the teacher received an automated welcome message with a template for communicating their problem (i.e., their demographics and overview of their issue) \update{along with a consent form}. Based on the teacher's reply, two \textit{facilitators} (first-author and a support organization personnel, 2 in Figure \ref{architecture}) forwarded the details to the nearest available \textit{caseworker} (3 in Figure\ref{architecture}) in their community (step 2, Figure \ref{workflow}). 

The caseworker then contacted the teacher on behalf of Saharaline and inquired about the teacher's well-being (step 3, Figure \ref{workflow}). Caseworkers belonged to the same community as teachers and worked for support organizations at their school. They were aware of the teacher's local context, such as the local language and their standing in the community, allowing them to capture the issues comprehensively. Being in the same community also allowed caseworkers to offer flexible interaction options (online, offline, inside or outside school) while accommodating teacher's busy schedules. The caseworker captured the teacher's well-being levels, work preferences, standing with school management and peers, and key problems (stressors) for which the teacher needed support. These details were recorded in a \textit{case file} (see supplementary material for an example) and sent to the facilitators via WhatsApp (step 4, Figure \ref{workflow}).

The facilitators then routed the case to an appropriate \textit{expert} (4 in Figure \ref{architecture}). For example, if the teacher's key stressor was teaching a classroom with varied learning levels post-pandemic, it was routed to the expert on pedagogical support (step 5, Figure \ref{workflow}). Other areas of expertise included occupational mental health, educational content, and educational technologies. When a case required multiple experts, it was routed to different experts in parallel. Experts worked remotely from their organization's headquarters. Based on the issue and the context provided by the caseworkers, experts formulated appropriate solutions, resources, and directions for the teacher, appending them to the case file (step 6, Figure \ref{workflow}). If experts had queries, they contacted the facilitators, who relayed them to caseworkers for clarification. The caseworker's responses were then provided to the experts. Over time, the case file accumulated all the issues and solutions for the teacher across multiple interactions.

Once experts added the solutions, facilitators informed the caseworker who scheduled a follow-up appointment with the teacher to share the solutions (step 7, Figure \ref{workflow}). A typical support cycle lasted seven days. This timeline allowed the caseworker to handle multiple cases in parallel while managing their regular work. To further provide longitudinal social support,  caseworkers reached out to the teacher a few days after providing the solution to gather feedback and inquire about any additional issues requiring support. Saharaline repeated the support cycle for several rounds if the teacher had follow-up questions or shared new problems.

\subsection{Pilot Deployment}
Saharaline was deployed for three months across three states in India (March to May 2022) in collaboration with four support organizations. Three organizations that engaged in Saharaline's design joined the deployment. Additionally, we advertised the study to other support organizations via a mailing list. Out of the five organizations that responded, we shortlisted one. This brought the total number of support organizations to four, which allowed us to keep the scope of the study manageable. \update{The final list of support organizations determined the schools in which we advertised Saharaline. We shortlisted six schools in communities where at least two of the four organizations had a presence.}
This incentivized support organizations to invest in teacher support alongside their regular activities. 

After receiving approval from the six schools, support organization personnel (with the first author) introduced Saharaline to the schools. Support personnel then advertised Saharaline on a bi-weekly basis in their teacher-focused WhatsApp groups for the duration of the deployment. Teachers responded by reaching out to Saharaline (as described in Section \ref{operation}), \update{during which they were provided with an online consent form that explained the data collection process in detail.} In total, 28 teachers reached out for support during the deployment. 

\subsubsection{Participants}
The six schools comprised 73 teachers. The first two schools were in a remote community in the state of Meghalaya. Both were government schools, \update{in the same block}, and catered to low-income communities. Out of the two support organizations working in these schools, one engaged with teachers and one with the higher management. The second two schools were low-income private schools in the \update{neighboring} semi-urban community of Gujarat. Out of the two support organizations working in these schools, one focused on students and the other on the higher management. The last two schools were \update{in two different urban districts in Karnataka}, where support organizations exclusively worked with teachers. All six schools were fully functional offline (i.e., in person). Out of the 28 teachers who used Saharaline, 12 taught primary school and 16 taught secondary school. 

We recruited four caseworkers across the six schools, with two caseworkers each in Karnataka and one each in Meghalaya and Gujarat. \nicki{The team, which comprised two women and two men, had the prior experience of working as} support personnel for different organizations for an average of two years prior to the deployment. They also lived close to the schools where they provided support. All five caseworkers were college graduates, with an average age of 26 years. 

We recruited five experts from three of the support organizations. They had an average age of 35 years and an average of 4.5 years of experience in  teacher training, designing educational content, developing educational technologies, and practicing occupational therapy. Three had master's degree while one had a doctoral degree. \nicki{{Three experts identified themselves as women while two as men}}. All were located in the cities where the organization's headquarters were located (i.e., not in school communities). Table \ref{tab:org-participants} presents the detailed demographics of all participants.

\begin{table}
 \center
 \renewcommand\arraystretch{1.3}
 \footnotesize
 \begin{tabular}[t]{|p{.75in}|p{2.35in}|}
 \hline
 \multicolumn{2}{|l|}{{\bf Teachers} (n=28)} \\ 
 \hline
  Gender & 
       \begin{tabular}{ll}
           Women: 18 & Men: 10\\
       \end{tabular}\\
\hline
 Age (years) & 
       \begin{tabular}{llll}
         Min: 25 & Max: 54 & Avg: 35.6 & S.D: 7.6\\
       \end{tabular}\\
\hline
Education (degree)  & 
       \begin{tabular}{lll}
           Bachelor's: 20  &  Master's: 8\\
       \end{tabular}\\
\hline
Region  & 
       \begin{tabular}{p{.55in}p{.46in}p{.58in}}
           Meghalaya: 8 & Gujarat: 9 & Karnataka: 11 \\
       \end{tabular}\\
\hline
Experience (years) & 
       \begin{tabular}{llll}
           Min: 3  & Max: 27  & Avg: 9.75 & S.D: 5.822 \\
       \end{tabular}\\

\hline
Phone use (years) & 
       \begin{tabular}{llll}
           Min: 4 & Max: 10 & Avg: 5.5  \\
       \end{tabular}\\
\hline
 Focus Subject & 
     \begin{tabular}{llll}
           Languages: 19 & Science: 9 & Math: 9 & Social Studies: 12 \\
       \end{tabular}\\
\hline
School Type & 
     \begin{tabular}{ll}
           Government: 16 & Private: 12 \\
       \end{tabular}\\
\hline
\hline
  \multicolumn{2}{|l|}{{\bf Saharaline Personnel} (n=11)} \\ 
 \hline
  Gender & 
       \begin{tabular}{ll}
           Women: 6 & Men: 5\\
       \end{tabular}\\
\hline
 Age (years) & 
       \begin{tabular}{llll}
           Min: 25 & Max: 42 & Avg: 31.8 & S.D.: 5.3\\
       \end{tabular}\\
\hline
Experience (years) & 
       \begin{tabular}{llll}
           Min: 3 & Max: 15 & Avg: 6.4 & S.D.: 3.7\\
       \end{tabular}\\
\hline
Role & 
       \begin{tabular}{lll}
           Caseworkers: 4 & Experts: 5 & Facilitators: 2\\
       \end{tabular}\\
\hline
Experts' Specialization & 
      Education Technologies: 1, Occupational Mental Health: 1, Teacher training: 2 , Educational content creation: 1 \\
\hline
Education (degree) & 
       
          \begin{tabular}{ll}
             Master's: 9  & Doctoral: 2\\
       \end{tabular}\\

\hline
\end{tabular}
\caption{Demographic details of teachers and support organization personnel involved in Saharaline intervention.}
 \label{tab:org-participants}
 \vspace{-6mm}
\end{table}

\subsubsection{Data Collection \& Analysis}
Our deployment produced three types of data that we analyzed to understand teachers' and support providers' experiences. Before collecting data, we obtained informed consent from all stakeholders. Our first dataset consisted of case files created for each of the 28 teachers who had contacted the helpline. The case files contained teachers' demographics, work preferences, a log of caseworkers' observations about teachers' issues, and any solutions provided (see supplementary materials for an example case file). 

The second dataset consisted of the interactions between teachers, caseworkers, and experts as part of the support effort. These included WhatsApp texts and audio messages that captured (1) initial messages that teachers reached out with when they shared their issues, (2) the caseworkers' interaction with the teachers through WhatsApp for coordination, clarification, and solution sharing and (3) interactions between caseworkers and experts for coordination and follow-up clarifications. In total, we analyzed 432 messages between the three stakeholder groups.

The third dataset came in the form of interviews that the first author conducted with the caseworkers (n=11), experts (n=15), and teachers (n=47) to capture their overall support-providing and seeking experiences during and after the deployment. \update{High-level topics for teachers included their motivation for seeking support, their challenges in getting the support, and their reasons for disengagements with the helpline. Topics for experts and caseworkers included their custom strategies to engage teachers, their challenges in working with resource constraints, and their collaboration strategies.}. In total, we obtained 75 hours of interviews and conversational audio recordings. Lastly, we also analyzed 21 pages of detailed notes captured by the first author while managing the helpline. Analyzing different forms of data allowed us to triangulate our findings \cite{creswell} and establish a strong sense of validity. 

We started our analysis by translating the data captured in local languages (Hindi, and Kannada) into English. Then, we engaged in qualitative coding using inductive thematic analysis \cite{terry2017thematic}. 
We began by taking multiple passes of all of our data to internalize the different accounts and perspectives. For  artifacts and messages, we began by going through the message logs, similar to \cite{varanasi2021Tag}, to conduct multiple rounds of open coding. Our unit of analysis was a single message sent by an individual participant. Even if the participant broke the message into multiple lines, we treated it as a single message. \update{All the researchers avoided any preconceived notions while constructing the codes and subsequent categories.} We followed a similar approach for the interview data. A second layer of internal validity was achieved via peer debriefing \cite{creswell} with multiple researchers, both external and internal to the study. Our analysis resulted in 77 codes (e.g., ``\textit{sources of tension}'', ``\textit{material assistance}''). The codes were then clustered into nine themes (e.g., ``\textit{Caseworker assistance in problem capture}'',``\textit{Contextual followups for teachers}'', ``\textit{Expert Clarifications}'').  

\subsection{Ethical Considerations}

\update{Our work took place in a fraught and sensitive context: low-income schools in India. 
Thus, we took steps to safeguard our participants, especially teachers. 
During onboarding, we sought consent both online, when teachers approached the helpline, and in-person, when caseworkers approached teachers. This ensured that teachers were aware of any risks associated with using the helpline. 
Protecting privacy and confidentiality was also of paramount importance in our deployment. All stakeholders, including caseworkers, facilitators, and experts engaged in a short course on (1) what privacy and confidentiality means, (2) handling different data types (e.g., personally identifiable data) and (3) best practices for providing support (e.g., how to ensure safe spaces for teachers to share their issues).}

\update{In addition, we took the confidentiality of the case file seriously. Teachers were assigned a pseudonym. Caseworkers were instructed to not record any form of personally identifiable information, such as school name, peers, and higher management names. Facilitators verified the case file before it was shared with experts. 
Caseworkers explained these practices to 
teachers when they initiated the first conversation with them. If a teacher sought multiple rounds of support, we ensured that the same caseworker was assigned to them. }

\section{Findings} \label{findings 5.1}
\begin{figure*}
 \centering
 \includegraphics[width=\linewidth]{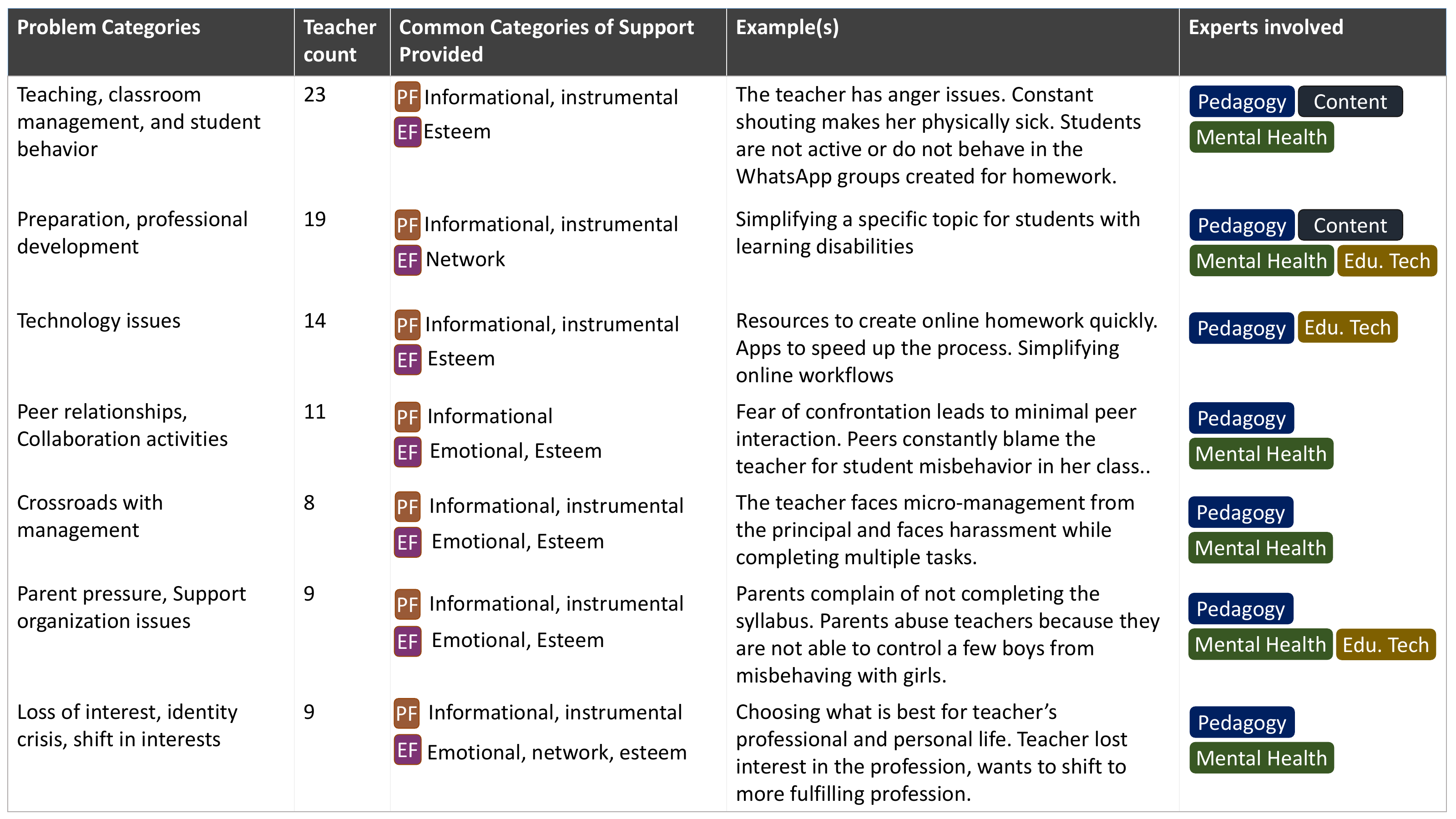}
  \caption{Table showing an overview of different types of problems and associated social support provided to the teachers. PF = Problem-focused, EF = Emotion-focused support}
  \Description{A graphic table showcasing a summary of different problem categories captured by saharaline. It also depicts teacher count, common categories of support provided along with prominent examples and experts involved.}
  \label{problems}
\end{figure*}
Saharaline was active for three months. During this period, 28 teachers reached out to the helpline and sought multiple rounds\footnote{One round is a full iteration of steps 1-7 as shown in Figure \ref{workflow}} of social support for different work stressors. This represented a significant proportion (30.12 \%) of the workforce in the six schools where Saharaline was promoted. At a high level, our findings show that Saharaline's collective social support approach effectively motivated teachers to share a wide array of problems impacting their occupational well-being. Figure \ref{problems} shows an overview of the problems that the teachers shared. They reflect the most common types of problems teachers in low-income schools faced in the Global South over the years \cite{cannanure2020, gavade2023, Singh_2023, bold2017}. Our findings also show that the collective efforts of the caseworkers and the experts to provide support were effective as teachers requested an average of three rounds of support with Saharaline.

In the following sections, we unpack our findings and discuss participants' experiences providing and receiving hybrid collective social support through Saharaline. Section \ref{4-1} sheds light on the conversations between teachers and caseworkers in the process of capturing teachers' problems. Section \ref{4-2} focuses on the co-production activities between the experts, caseworkers, and facilitators as they worked together to solve teachers' problems. Finally, Section \ref{4-3} focuses on the caseworkers' efforts in delivering solutions to the teacher, and the teachers' subsequent feedback. All participant names are replaced with pseudonyms to ensure anonymity.

\subsection{Capturing Stressors: Problem Stories \& Scenarios}
\label{4-1}

\subsubsection{Teachers' Motivations for Seeking Support}
\begin{figure*}
 \centering
 \includegraphics[width=\linewidth]{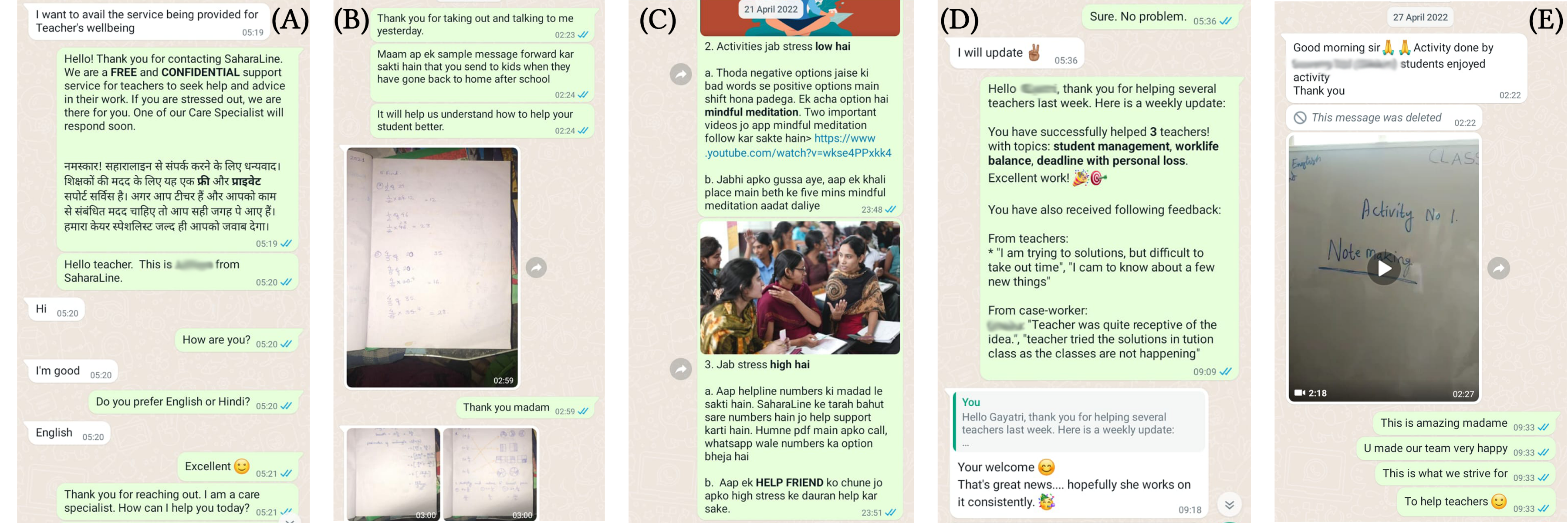}
  \caption{Message excerpts from helpline; (A) Automated greeting message when teacher reached out to Saharaline; (B) Caseworkers requesting teacher for additional context for her problems by asking her to share sample home work she shares with the students after school \& teacher sharing materials in response; (C) An example of how caseworker localizes the expert solutions from the case file. The first snippet introduces the teacher to mindful meditation techniques and the second snippet shares important helpline numbers and introduces the notion of a ``help friend''; (D) Curated summary for experts. Message consisted weekly highlights, teacher and caseworker's feedback; (E) A teacher sharing a video of her executing suggested solutions in the classroom by the expert. }
  \Description{Five sample WhatsApp message excerpts shown from the helpline. First image is the automated greeting message when teacher reached out to Saharaline. Second message depicts caseworkers requesting teacher for additional context for her problems. Third image shows an example of how caseworker localizes the expert solutions from the case file. Fourth images depicts curated summary for experts consisting weekly highlights, teacher and caseworker's feedback. Final image shows a teacher sharing a video of her executing suggested solutions in the classroom by the expert.}
  \label{whatsapp-screens}
\end{figure*}

Two types of teachers contacted Saharaline. The first type (n=17) were \textit{explicitly aware} of the nature of their problems and could narrate them by identifying the source of their stressors. They could also describe the impact the stressors had on their work lives. By sharing their narratives, they sought targeted solutions for their stressors. Common problems expressed by explicitly aware teachers included impending deadlines (e.g., syllabus completions), stressful events (e.g., exams), classroom management, preparation, and teaching assistance. One example conversation\footnote{Emoticons are omitted from chats to focus on readability} between Gita, a social studies teacher, and Saharaline is presented below:

\begin{myquote}
\textbf{4:41 PM. Gita}: \textit{Hello. Are you there? I am Gita} \newline
\textbf{4:41 PM. Saharaline}: [Automated welcome message (See Figure \ref{whatsapp-screens}.A)] \newline
\textbf{4:50 PM. Representative}: \textit{ Hello, Gita teacher. Thank you for reaching out to Saharaline} \newline
\textbf{4:50 PM. Representative}: [Template message asking for demographic details and the issue] \newline
\textbf{5:29 PM. Gita}: [Omitted Demographic details]\newline
\textbf{5:29 PM. Gita}: \textit{Do you help with parent issues? I have a P.T.A. meeting in two weeks and the parents for a few students are not happy.} \newline
\textbf{5:32 PM. Gita}: \textit{It is becoming very stressful thinking about how to handle them.}
\end{myquote}

Like Gita, teachers in this category found it easier to establish trust, clearly narrate problems, and request assistance. 

The second type of teachers (n=11) that contacted Saharaline were \textit{implicitly aware} of their problems, but found it difficult to identify their stressors and narrate how the stressors were impacting their work lives. They contacted Saharaline to express their curiosity about the helpline, share their lived experiences in schools, and ask broad questions to make sense of their work lives. Common problems within this category focused on emotional distress experienced at work, subsequent fatigue and burnout, issues with their professional image, and the impact of personal struggles on work. Less common issues included relationship complications with higher management and peers. Prerna, a Hindi teacher who reached out to Saharaline, shared the following in a post-support interaction:
\begin{myquote}
    ``\textit{When I first reached out to your helpline, I was not even sure why I was messaging you or what I wanted to talk about. At that time, I was just feeling mentally tired and just the thought of going to school was making me want to run away \dots I still remember, I just messaged because the helpline for teachers made me curious. I never heard of something like this [Saharaline] \dots Later when I was answering Asmeen's [caseworker] questions I realized that the management was giving unrealistic amount [of work] \dots I was also just not interested in doing teaching anymore \dots But I thought it was my mistake and I was just feeling guilty about it.}''
\end{myquote}

Teachers in this category also found it challenging to articulate their problems with their peer networks. Saharaline, being an independent intervention outside their school ecosystem, provided a viable alternative to seek support. Unlike explicitly aware teachers, they did not have clear expectations and agendas when approaching Saharaline. As a result, these teachers shared abstract narratives, which meant that caseworkers had to put in more effort to surface and capture concrete problems for Saharaline to assist.

\subsubsection{Validating \& Capturing Problem Stories}
Regardless of the teacher's motivation type, caseworkers contacted the teacher and captured their problems. To lower the threshold to share problems, caseworkers encouraged teachers to start with their recent lived experiences. Caseworkers referred to these narratives as teacher \textit{stories}.  
 
To surface these stories, caseworkers shared how they had to plan the first interaction with the teacher carefully. To create an empathetic and positive first impression, they carefully considered the medium of communication (WhatsApp message/call vs. in-person conversation), time of communication (school hours/breaks vs. after-school), environment (inside school vs. outside school), and motivation of the teacher (explicitly vs. implicitly aware). Minu, one of the caseworkers, shared some of her considerations while deciding what medium of communication she had to use when she interacted with teachers for the first time:  
\begin{myquote}
    \textit{I have actually worked with teachers from these communities before. Teachers are quite apprehensive when approached through WhatsApp or any other technological platform. If they are apprehensive, how will they open up? Moreover, interaction via online platform does not help us know what emotional state they are in that day \dots see their facial expressions \dots. My first long interaction is always offline to observe all of this while I try to find their problems \dots we can easily connect this way. But I never force them. Ultimately, they choose.}
\end{myquote}.

Generally, our analysis revealed that caseworkers who showed more flexibility while engaging with teachers received more engagement from teachers. 

\subsubsection{Translating Problem Stories to Problem Scenarios} 
The stories shared by teachers were extremely rich in detail. Caseworkers found it ineffective to share them in their original state with Saharaline experts for solutions. Instead, the caseworkers used different nudges to encourage teachers to reflect on their stories, extract essential information, and create concrete {scenarios}, which they then could register in the case file for experts. 

The caseworkers used scoping questions as a common technique to narrow down the exact stressors in the teachers' stories. For instance, when caseworker Abdul met Ragini, a secondary school Science teacher, her story revolved around how her class students had become extremely unruly and unmanageable post-pandemic, because of which she was not able to complete her curriculum, contributing to stress. While it was clear to Abdul that the teacher was struggling with classroom management, he still reached out to her on WhatsApp and asked follow-up questions to narrow down the specific aspects of management she was struggling with:

\begin{myquote}
    \textbf{10:31 AM. Abdul:} \textit{Ragini Maam,  yesterday, you told me you struggle with controlling children in the classroom. Once the period starts, between when you step into the class and when you come out,  what times are the most difficult?}\newline
     \textbf{10:33 AM. Abdul:} \textit{Your answer will help me share exact issues with the expert. If you are busy,  please leave a message} \newline
    \textbf{10:33 AM. Ragini:} \textit{The main problem is making them sit quietly and behave. That itself does not happen} \newline
     \textbf{10:39 AM. Ragini:} [Voice message in which Ragini shared an example of how students talk among each other and how she took 20 minutes to silence the class and start the lesson]\newline
     \textit{[Abdul asked a few follow-up questions]}
\end{myquote}

Based on the conversation, Abdul created a concrete problem scenario indicating that Ragini struggled with classroom management strategies, particularly in setting clear expectations and modeling ideal behavior.  
Another strategy used by caseworkers was to nudge teachers to consider the relative impact of the problems on their work lives. Caseworkers shared how some teachers reported several problems at once in their stories. To effectively support them, the caseworkers assisted the teachers in understanding, ranking, and prioritizing problems that affected their well-being the most. Based on teachers' responses, caseworkers broke down the story into multiple scenarios and prioritized the scenarios.

Caseworkers also nudged teachers to provide additional contextual information to their stories that could uniquely anchor their problems to their local settings. For example, caseworkers requested supporting materials to make the scenario as concrete as possible. Our analysis revealed that caseworkers included all kinds of supporting materials, including documents like book chapters, the curriculum schedule of the school, and detailed descriptions of students present in teachers' narratives. They complemented this effort by including passive indicators they observed in person while listening to the teacher's problems, such as environmental, social, and infrastructural details.

For instance, Harpreet, a caseworker, was assisting Parnita, who was teaching Mathematics to secondary school students in a remote school. Post-pandemic, their school had adopted a hybrid approach to teaching. Due to pressure to complete the syllabus, Parnita was instructed to teach half of the curriculum at school and send notes for the other half via WhatsApp to students' parents. However, there were three students whose families did not have phones or Internet access. The teacher was extremely stressed and reached out to the helpline. While producing tangible problem scenarios, Harpreet included in-depth details of the three students, including their parents' occupations, other fellow students' families in the vicinity, pictures of student neighborhoods, and internet cafe and photocopy shop availability. Esha, a teacher trainer and a pedagogical expert, shared how these details were invaluable in designing peer-based activities by including the student's peers who lived near his house.

In another example, caseworker Minu shared how one of the teachers needed content and pedagogical tips for a specific chapter. To provide context for the expert, she took a photocopy of the book chapter,  annotated the issues across the margins, took a photo, and shared it with the helpline.
The caseworker then submitted the problem scenarios to the helpline by recording them in the case file. In instances when their support overlapped with multiple teachers, caseworkers optimized their time to submit the information by sharing photos of handwritten notes or voice notes on WhatsApp with the helpline facilitators who transcribed, formatted, and appended the scenarios to the case file.  

\subsubsection{Caseworker's Challenges In Capturing Problem Stories and Creating Problem Scenarios} \label{5-1-Challenges}
Crafting problem scenarios that encapsulated teachers' stressors was complicated for caseworkers despite their familiarity with teachers' work environments. For instance, caseworkers found it more difficult to surface stories from teachers who were implicitly aware of their problems. All caseworkers unanimously shared how they had to spend more time with the implicitly aware teachers. They had to engage in alternative strategies that allowed them to obtain the stressful themes in the teachers' lives organically. Harpreet shared how her first interaction with a teacher who contacted the helpline out of curiosity was to meet and initiate a general conversation about herself. These unrelated and open-ended conversations allowed the teacher to open up and share their stories. Harpreet shared: 
\begin{myquote}
   ``\textit{Once he opened up with his stories, it was much easier for me to ask follow-up questions and understand where he was stressed. [This] teacher was mentally burned out but did not come to terms with it \dots his stories gave away those issues. I suddenly came up with this idea. I told him that we at the helpline are sharing 10-min video prompts that are around different work topics, teachers just see them and I come after a few days and we talk about it \dots helpline [facilitator] said they could do something like this with the experts \dots Mental health for teachers was one topic \dots Guess what topic the teacher chose? Mental health.''} 
\end{myquote}

The idea of such small prompts was effective with implicitly aware teachers as they acted as catalysts for teachers to open up for more conversations with the caseworkers. However, in a few instances (n=3 who needed emotion-focused support), caseworkers asked the helpline facilitators to connect them directly with the experts. When the caseworkers went back to interact with the teachers, they put the expert on the call so they could ask focused questions. Answers to those questions helped caseworkers shape scenarios they found difficult to do independently.

Another challenge for caseworkers in having meaningful conversations was working around teachers' lack of personal space and time in the work environment. All low-income schools that the caseworkers visited had limited real estate for classrooms and staff rooms. It was common for teachers to shuffle between an empty classroom and an unoccupied staff room as the rooms saw constant movement from students and teachers. Moreover, when other teachers saw a particular teacher engaging with the caseworkers in serious conversations, they gathered around the teacher and steered the conversation away from the particular teacher's concerns. This practice complicated the caseworker's effort to create scenarios mapped to the concerns of the teacher who reached out to Saharaline for support. To overcome this issue, several teachers (n=11) requested caseworkers to meet outside their school, such as makeshift playgrounds in front of schools and public spaces near schools, to maintain privacy. 


\subsection{Clarifications, Development, \& Refinements: Problem Scenarios to Solution Scenarios}
\label{4-2}

\subsubsection{Experts' Clarifications to Produce Solution Scenarios} \label{5-2-Clarifications}
Once the caseworkers added the problem scenarios and relevant supporting information to the case file, the facilitators assigned them to the experts. Based on the stressors mentioned in the problem scenario, a single teacher was mapped to one or more experts in parallel. After reading the case file, if experts needed clarifications and additional context before developing solutions, they were encouraged to interact with the caseworker through the helpline. 

Our findings suggest that experts appreciated this layer of separation for a few reasons. Experts shared that, by choosing to distance themselves from the process of problem formulation and context development, they could devote more time to finding meaningful solutions for teachers' problems. This allowed experts to scale their support by working on multiple cases simultaneously. Pedagogical expert Esha shared:

\begin{myquote}
  \textit{I like to have some emotional distance from the teacher for whom I am providing a solution while having a way to get more information. Being remotely available through this kind of app service is perfect \dots Don't get me wrong. The reason I say this is because my job is not just to provide a solution that reduces stress for that one teacher, but to do justice to all teachers equally. Abdul's [caseworker's] job is a really satisfactory one, but if I am doing that too and engaging with teachers \dots teachers' lives are so challenging and stressful that I cannot move beyond that first teacher.} 
\end{myquote}

When needed, experts reached out to the helpline for additional clarifications. In total, we recorded 128 requests for clarifications raised by experts. On average, this amounted to two clarifications per solution provided to a teacher. Within these, an essential type of clarification sought by experts was technical specifications related to their field of specialization. Common examples included teachers' pedagogical strategy (e.g., student-centered learning), curriculum details (e.g., national or state curriculum), professional development certification details (e.g., state-sponsored or support organization sponsored), and technological configurations (e.g., availability and the resolution of the projector).

Another type of clarification sought by experts pertained to potential conflicts they anticipated their solutions could introduce into teachers' work lives. They shared three distinct scenarios. One possibility was that their solutions could conflict with the school-specific norms imposed by school management. Another concern was that their solution could conflict with other problems faced by the teacher that were not yet added to the case file. Lastly, they were also concerned that their solutions might conflict with those provided by other experts. 

In all situations, the experts requested additional information to ensure their solutions did not create further distress when teachers implemented them. One such instance occurred when Sunita, a Science teacher, contacted the helpline to express her stress stemming from interactions with students' aggressive parents. In the problem scenario, the caseworker explained that due to a significant decline in students' learning levels following the pandemic, the school had directed teachers to revisit the previous year's syllabus. This decision was met with resistance from parents, who harassed and accused the teacher via WhatsApp of not letting their children progress. Sunita was understandably distressed, as she felt she was doing everything she could to educate the students. In response to this problem scenario, Jayanth, the pedagogical expert assigned to this case, wanted to share activities that Sunita could distribute to parents via WhatsApp to alleviate some of her pressure. However, Jayanth sought clarifiation before providing the solution as their solution conflicted with that of Priti, the occupational mental health expert:  
\begin{myquote}
\textbf{9:23 AM. Jayanth}: \textit{Hi.... I just saw Priti ma'am's [mental health expert] solutions. There is a slight issue.} \newline
\textbf{9:24 AM. Jayanth}: \textit{She thinks the parent's harassment is causing stress to Sunita [teacher]. She mentioned some strategies to help her handle them, which is good...... }\newline
\textbf{9.26 AM. Jayanth}: \textit{One suggestion she shared is to stop interaction for a few weeks and ask the principal to intervene till she feels better.    }\newline
\textbf{9.28 AM. Jayanth}: \textit{.... I designed a few exercises for Sunita to involve parents in their children's learning for the syllabus they wanted the teacher to teach. My goal was to help teacher give some homework that could be done by parents and students together. }\newline
\textbf{9.30 AM. Jayanth}: \textit{But I don't want to unnecessarily confuse the teacher by giving contradictory instructions. Can you confirm with Priti ma'am and let me know what she thinks?  } \newline
\textbf{9.39 AM. Jayanth}: \textit{If she thinks its not okay..... I will probably add a note or just come up with something else... I have left comments for her on the case file for clarity}
\end{myquote}

\subsubsection{Refining the Solution Scenarios}
Once the expert received clarifications, they developed a rough draft of the solution and added it to the case file. Experts called these \textit{solution scenarios}, mirroring caseworkers' use of problem scenarios. A typical draft of a solution scenario comprised three parts. The first contained the reasoning behind the teacher's problem (e.g., ``\textit{you [teacher] are facing this fatigue because you are not able to say \textit{NO} to the management certain situations}''), followed by contextualized instructions for the teacher (e.g., ``\textit{explain to the teacher that improving learning levels of students will take time}''), and ending with supporting audiovisual materials that showed the implementation of the suggested instructions in a contextual scenario (e.g., hand-drawn image of a leader board system for high school students to encourage increased productivity and thereby learning levels). After adding a draft solution scenario, the expert sought another round of clarifications from caseworkers to refine and finalize their solutions.

Caseworkers used this opportunity to share their perspectives on the practicality, the resource requirement, and any additional burden the proposed solutions could create on teachers. In a similar issue with parents for another teacher, Romila, another expert, suggested the teacher document her work with the students as proof to show the parents. The caseworker pushed back against the suggestion by reasoning that written documents could add to the teacher's current documentation burden in school. Instead, the caseworker suggested the creation of one minute video logs (similar to YouTube Shorts) of their classes to achieve the same objective with less work. 

Experts also sought explanations to refine their solution scenarios based on the caseworker's understanding of whether the teachers had tried prior strategies to address their stressors. To confirm this, they requested the caseworkers to provide details of any steps the teachers took as part of their efforts. These requests comprised the biggest proportion (approx. 38\%) of the overall clarifications sought by the experts. Based on this feedback, experts refined their solutions and proposed more effective strategies. The high frequency of these clarifications motivated caseworkers to include this information as contextual information within their problem scenarios. 

In total, we found that experts created 79 solution scenarios throughout the deployment period, comprising five different types of social support (see Figure \ref{problems}). The most common and least common types of support provided were informational support, a type of problem-focused support \cite{Carroll_2020} where information is provided on ways to reduce stress (72\%), and network support, a type of emotion-focused support where network connections are provided to manage the emotions, rather than altering the situation (6\%). In the finalized solution scenarios, the experts also added instructions for caseworkers to share the solutions with the teachers. These explanations added a layer of assistance to the instructions meant for teachers, especially for those solution types that the experts thought were unfamiliar or tricky to follow. Moreover, when experts realized that the teacher’s strategies for problems were ineffective and detrimental to their well-being, they also instructed the caseworkers on how to inform teachers about it. For example, the mental health expert Priti, shared the following instructions to the caseworker in the teacher's solution scenario:

\begin{myquote}
\textit{``The teacher's idea regarding partitioning the dull students and giving them more importance is a good idea but please make her understand (when she is alone) that doing it in front of the bright students might create a feeling of resentment/dislike towards the teacher and [the students might] become disinterested with the subject. It will unfortunately only add to the problem for which the teacher reached out in the first place and make her more stressed.''}
\end{myquote}

To help experts add their solution scenarios to the case file on time, the facilitators provided recurring alerts and periodic reminders of clearly articulated deadlines based on the caseworker’s subsequent scheduled visit to the school. When the experts required more time, they contacted the facilitators to keep them informed. In those cases, caseworkers had to reschedule their appointments with the teacher accordingly. In the entire study, there were four such instances where experts requested extensions.

\subsection{Delivering Solutions \& Follow-ups: Solution Stories}
\label{4-3}

\subsubsection{Strategies and Challenges}
Once experts finalized the solution scenarios, caseworkers shared them back with teachers. However, caseworkers felt the solution scenarios in their original form were dense in information. Consequently, all the caseworkers simplified the scenarios to increase the teachers' understandability and the subsequent likelihood of implementing the solution in their work lives.

One simplification strategy was to familiarize themselves with the solution scenario and express them as a series of illustrated snippets through WhatsApp (see Figure \ref{whatsapp-screens}. C). Caseworkers referred to the snippets as ``\textit{solution stories}''. They sent these stories a few hours before interacting with the teacher to invoke curiosity towards the solution. The story illustrations were either sourced from the supplementary materials in the solution scenarios or curated separately through online searches to localize the solutions to teacher's work settings.

In some cases, caseworkers took an additional step by demonstrating to teachers how to implement the solution stories, with the hope of invoking discussion. To do this, caseworkers either met the teacher in person or recorded themselves demonstrating the activity. For instance, Suman, a secondary school Science teacher, contacted Saharaline about class management issues. She described how the caseworker, Asmeen, visited and demonstrated strategies: 

\begin{myquote}
    \textit{``Asmeen was very patient \dots After school, we sat in a classroom and discussed my issues. She has seen me teach. So we imagined that the classroom had students and he showed me some management techniques. I am actually a new teacher and she is familiar with my students. So after she showed me I asked her, ``what about [student name]. He will get too excited and disrupt the class?!'' and then she replied what to do in this new situation \dots This exchange was helpful \dots One time I had to record her solutions because I was in hurry \dots I am so tired at the end of the day. It is morally uplifting to see someone sitting with you and then working through your problems.  ''}
\end{myquote}

Through these strategies, caseworkers aimed to make solution scenarios easily understandable for teachers. The demonstrations encouraged teachers to implement the solutions by reducing barriers, such as hesitation or embarrassment. Caseworkers' physical presence also encouraged teachers to actively resolve any follow-up questions.

Despite the strategies, caseworkers experienced several challenges while delivering solutions to the teachers. A key challenge was convincing teachers to implement emotion-focused solutions to manage the negative emotions associated with stress rather than altering the stressful situation or eliminating the stressor. For instance, Sofia, a young teacher with six years of teaching experience, reached out to the helpline with issues of burnout at work, loss of interest in teaching, and a desire to discontinue the job. She felt like she was ``\textit{stuck and could not do anything about it}''. The caseworker, Abdul, was responsible for sharing self-care activities, such as recognizing negative thoughts about her job and trying to transform them into positive ones (referred to as cognitive restructuring \cite{Clark_2013}). It took multiple visits by Harpreet to persuade the teacher that these activities would have long-term benefits.

Another challenge in delivering solutions was that it took around seven days to provide a complete round of social support (steps 1-7 in Figure \ref{workflow}). By the time caseworkers returned with solutions, teachers had implemented their own solutions on a few occasions (n=4). In such cases, case workers had to report these new solutions to Saharaline and confirm the applicability of the expert solutions in light of the teacher's own solutions. If the expert suggested changes, the updated solutions were shared with the teacher. On certain occasions, teacher's own support efforts contributed to additional stress rather than relief. In those situations, caseworkers had to make a substantial effort in convincing teachers, especially with seniority, to discontinue such practices. For example, two teachers reached out to Saharaline, sharing how they were experiencing harassment from senior teachers in the school. As a coping strategy, they stopped engaging with all of their peers, including essential activities such as teacher development training and parent-teacher meetings. Caseworkers had to first work with the teachers to help them recognize and stop their detrimental coping mechanisms before they could suggest new activities proposed by the expert.

On rare occasions, caseworkers, due to their neutral position in the school ecosystem, became an active part of the solution for problems involving power dynamics or persistent systemic issues. In such instances, experts encouraged caseworkers to strategically approach management and increase the visibility of systemic issues on behalf of teachers. Caseworkers shared how they had to conduct these conversations carefully, keeping teachers' identities anonymous. 

\subsubsection{Factors Determining Teacher's Solution Adoption \& Follow-up Support} \label{5-3-solution-adoption}
Our data show that teachers, on average, availed three rounds of support (min= 1, max = 7) from the helpline. The subsequent rounds of support were either a follow-up of the same problem or a new problem. While the data suggest that the teachers found Saharaline to be beneficial for longitudinal support, it is important to understand what factors teachers considered to implement a solution and even request follow-up support.

An essential factor was the priority of the problem in the teacher's life. Saharaline support took anywhere between four and seven days to provide a solution to the teacher's problem. On occasion, between the time when the teacher shared their problem and when the caseworkers shared the solutions, other higher-priority problems took precedence. On such occasions, the teacher either put the implementation of the suggested solution on hold or used the follow-up visits to address the higher priority problem. A teacher shared one such example:

\begin{myquote}
    \textit{`` Yes, Asmeen [caseworker] had shared the solution for the new [student-centered] teaching method that the [principal] sir was asking from me. We had one conversation about what to do for my sandhi-vicchhed [a grammar topic in Hindi] chapter. But I could not implement them because exams were scheduled a week earlier all of the sudden. I had a completely different problem then. I needed help in tackling students with different learning skills who were not able to grasp the concept and I was stressed about completing the syllabus. I requested Asmeen to help me with some techniques in completing the hard to grasp lessons to finish the syllabus that is appearing in this exam.''}
\end{myquote}

Another factor teachers considered while incorporating the solution was the perceived amount of effort necessary to implement the solutions and the time required to see the results. Teachers were more likely to try a solution if the caseworkers broke it down into comprehensible pieces and clearly demonstrated its feasibility. 

The likelihood of follow-up questions was also determined by perceptions around how much effort would be required to incorporate solutions. When teachers found it easy to implement the solutions in their work lives and perceived the results as effective, their trust in Saharaline increased, and they reached out for subsequent assistance. We also saw the likelihood decrease when the caseworker shared solutions in a delayed manner. Among the teachers who took subsequent rounds of support from Saharaline, eight teachers requested additional support for the same issue, four teachers shifted to a new problem, and fourteen requested for both. Interestingly, caseworkers also initiated a few follow-ups in certain circumstances when they felt that the teachers were showing symptoms of stress, such as melancholic status messages on WhatsApp or when they heard distress updates from other peers. 

During the follow-ups, caseworkers also sought feedback from teachers on the provided solutions and shared it with the facilitators. \update{ Interestingly, facilitators revealed how they realized that the initial Saharaline's design lacked appropriate structures to inform experts about the teachers' experiences for whom they provided support. To address this issue, facilitators devised the idea of gamified weekly summary prompts, which they shared on WhatsApp. These prompts contained updates on delivered solutions}, such as the count and category of solutions the expert shared the previous week and a sentence summary of the feedback from teachers and caseworkers (see Figure \ref{whatsapp-screens}.D).  

\subsubsection{Perception of Saharaline}
We now briefly discuss the technological experiences of caseworkers, experts, teachers, and facilitators. All stakeholders unanimously shared that the choice of everyday technologies like WhatsApp was helpful for easy adoption and seamless use of the platform in receiving and providing social support. This was a key factor because all the stakeholders were on a tight work schedule during their work hours, and they greatly appreciated not having to learn a new technology. Pushpa, one of the experts, shared her perspective on this matter:

\begin{myquote}
    \textit{``One good thing about using WhatsApp was that it was easy to go back and forth because sometimes a question suddenly comes to mind while working on the solution or cooking. It is easy to go back and check: will this work for that teacher? Regarding accessibility, I felt a lot more comfortable and I am glad it was on WhatsApp and not on a new app I have to learn. I already use so many custom apps for this and that. \dots''}
\end{myquote}

However, a few teachers also shared (n = 5) how interactions on WhatsApp added to the already overwhelming number of messages they received. These teachers indicated a strong preference for in-person interactions. By contrast, experts, who were comfortable with technology use, requested an option to seek additional review from another expert as a second pair of eyes to increase the veracity of their solutions. Lastly, they shared how, as teachers requested more follow-up support sessions, it took substantial work to cross-check older WhatsApp clarifications with the case file and make connections. Instead, they preferred a copy of their message interactions added to the case file to maintain prior context of the problem in the same file where they provided the solutions.

\section{Designing for Improved Collective Social Support}
Our findings show that Saharaline surfaced a wide range of teacher issues and support needs. Here, we reflect on our findings, discussing the benefits and tensions of Saharaline's highly versatile approach to delivering collective social support (Section \ref{5-1}). We also highlight how Saharaline's design enabled decentralized production of knowledge, which improved the overall distribution of expertise and resources (Section \ref{5-2}). Finally, we reflect on our choice to deliver collective social support via WhatsApp, itself often a stressor for teachers (Section \ref{5-3}). 

\subsection{Versatility of Collective Social Support Practices} \label{5-1}
In low-resource settings, disparities in school infrastructure and resources lead to significant differences in teachers' local work practices \cite{cannanure2020, Alcott_Rose_2015} (e.g., the curriculum they follow, language of instruction, available infrastructure, and access to professional development). Saharaline caseworkers effectively captured these local practices, along with the teachers' problems, particularly benefiting implicitly aware teachers who struggled to articulate their issues. Experts then combined this local information with their cross-cultural knowledge, which included domain expertise and deep understanding of the Indian education system. This collaborative effort produced contextual solutions that caseworkers and teachers could readily implement in local contexts.

Collective social support was effective due to this inter-dependency between localization practices (by caseworkers) and theory-driven practices (by experts) to provide balanced support. Caseworkers and experts carefully negotiated their inter-dependencies for different types of support. When Saharaline provided problem-focused social support, the team's inter-dependencies were well-defined, relatively simple, and mainly carried out on WhatsApp. \update{These observed patterns align with problem-focused support in collaborative care environments, akin to monitoring health measurements through simple technologies and straightforward protocols \cite{caldeira2017}. The utilization of uncomplicated technologies facilitated ease of learning, \nicki{while clear-cut activities promoted improved coordination among stakeholders, reaffirming \citet{amy2020}'s findings}. In our study, the majority of problem-focused stressors, encompassing teaching, preparation, and administrative issues, manifested as structured problems with systematic solutions, thereby enabling effective collective social support.}

By contrast, when teachers required emotion-focused support, the inter-dependency between actors increased in complexity, requiring carefully coordinated collaborations delivered in a hybrid manner. This was because these issues were often implicit and less structured. \update{Collective support systems implemented in resource-constrained environments face challenges in addressing such implicit support requirements. These systems often depend solely on the collaborative efforts of individuals within the community (e.g., parents seeking relatives for educational support \cite{Madaio20}) or external experts who are not part of the community (e.g., researchers providing expertise in resolving technological issues \cite{PetersKuria_2014}). This reliance on either community collaboration or external expertise makes it difficult to fully comprehend implicit support needs. In our study, we identified a middle ground by empowering caseworkers, who possessed local awareness, to bridge the gap and connect with experts.} early in their communication with teachers to better capture these implicit requirements and improve the problem scenarios they developed (Section \ref{5-1-Challenges}). Similarly, experts requested clarifications from caseworkers when suggesting solutions (Section \ref{5-2-Clarifications}). \nicki{This versatility in inter-dependency enabled Saharaline to effectively handle a wide range of both known and new issues, while keeping the burden for the system and individual roles low \cite{Archer_2012}}.   

However, the versatility of the support may have come at the cost of the time required to provide it. Our findings show that it took an average of six days for the Saharaline team to provide support. Caseworkers consumed most of this time (roughly four days) to establish connections with teachers, build rapport, and capture their local experiences and problems. Experts took roughly two days to craft solutions, and facilitators took a few hours to coordinate the overall process. In a few time-sensitive cases (n=3), by the time the Saharaline team delivered solutions, new, higher-priority problems had surfaced in teachers' lives (Section {\ref{5-3-solution-adoption}). In such situations, teachers put the original problem and associated solutions on hold, and either picked them up later or never tried them.

Another challenge was that Saharaline relied heavily on the role of caseworkers, who were the key stakeholder that interfaced with the teachers. Success of personalized case management systems rely on these actors who can build rapport, and encourage reflection and discourse \cite{Tseng_2022}. This study adds to this criteria wherein caseworkers also carefully separated their role of problem translators (``emotion'') from the experts (``technical'') who solved the problem by establishing clear boundaries, something that prior research found challenging \cite{Tseng_2022}. \nicki{But this also means that the intervention relied on low-income schools having caseworkers already deployed. This might not be always the case with all low-income schools. In such cases, support organizations need to play a critical role in helping schools train caseworkers from the local community that occupy a similar neutral position. More research is needed to understand how such interventions can be deployed and sustained in low-income schools that lack easy access to support organizations and their deployed roles}.

These findings highlight a tension in providing collective social support to teachers. On one hand, receiving personalized support and expert-crafted solutions from domain experts within a week (i.e., six days) may be considered a fast turnaround for such a service. Indeed, our goal in designing Saharaline was to deliver longitudinal support to teachers over long periods of time, not to run an emergency service. On the other hand, some teachers' priorities changed before Saharaline could provide support. Given this, a logical solution may be to try and provide support more rapidly. However, we would caution against the assumption that faster support would be better. Taking time to build rapport and deeply understand teachers' contexts, and carefully considering potential repercussions of any provided solutions on teachers' lives, are important steps that necessarily take time. \nicki{As such, ``fast-tracked'' solutions may run the risk of harming the target populations and exacerbating their problems \cite{Madaio20}}. 
Instead, researchers deploying such collective social support in the future should carefully consider if workers will need time-sensitive or urgent support and plan accordingly. In Saharaline's case, most teachers who reached out for support did so for non-time-sensitive problems, and appreciated Saharaline's approach to providing personalized, localized assistance via multiple rounds of support. 

\subsection{De-centralized Knowledge Production} \label{5-2}

Saharaline's versatile design involved experts from multiple support organizations, thereby enabling decentralized knowledge production that was culturally and socially diversified. \update{Such diversified, decentralized support systems function as a robust source of alternative knowledge for teachers \cite{pendse2022,Mignolo_Walsh_2018}. They play a crucial role in raising awareness and addressing sensitive topics, (e.g., mental health), which might be considered taboo and not addressed by centralized support systems governed by management. Here, we further unpack this decentralized form of knowledge production.}

When teachers contacted the helpline, they shared stories containing tacit knowledge gained over years of teaching. This knowledge encompassed implicit practices related to classroom and community culture, student behaviors, and commonly encountered challenges. \citet{Shulman_1986} defined such forms of applied tacit knowledge as pedagogical content knowledge. Caseworkers then translated these stories into explicit, declarative knowledge scenarios, incorporating language and references that reflected the technical background and skills of the expert. In documenting scenarios, caseworkers assumed the role of active knowledge translators, systematically organizing scenarios and adding context.


The experts, in turn, analyzed the problem scenarios using their conceptual knowledge of mental health, pedagogy, and educational technology. They used this knowledge to evaluate the problem severity and generate recommendations in the form of solution scenarios. These contained solutions for teachers alongside, more importantly, instructions for caseworkers in the form of procedural knowledge. As such, experts served as \textit{knowledge providers}. Caseworkers then again assumed the role of knowledge translators, taking the relevant instructions from the expert's solution scenarios and localizing them to suit the teacher. As translators, their task involved simplifying concepts into a language comfortable for the teacher within their community. Decentralization of roles also facilitated knowledge production at multiple steps involving multiple roles. Saharaline's design successfully distributed these knowledge production efforts, enabling the system to be approachable for teachers, \update{while being managable to carry out by stakeholders}.   

For their part, facilitators played an essential role in ensuring the timeliness of knowledge production and the seamless transfer between caseworkers, experts, and teachers. These activities can be categorized into two types of tasks. The first are \textit{coordination} tasks that facilitators engaged in to ensure the helpline was functioning as expected (e.g.,connecting teachers with caseworkers, routing clarifications between experts and caseworkers, and rescheduling interactions between caseworkers and teachers when an expert's solutions was delayed). These tasks also involved translating between different languages (Hindi, Kannada for caseworkers and English for experts) to enable conversations. The second type of task revolved around \textit{information management}, which encompassed information collection, organization, and communication at various levels. These tasks included capturing caseworkers' audio and textual notes in the case file, mapping appropriate experts to the case file based on the problems shared, weekly information synthesis of details of the support provided, and sharing gamified prompts with the experts.  

Facilitators' efforts, while simpler than knowledge production tasks undertaken by caseworkers and experts, consumed a lot of time and energy to ensure smooth functioning of Saharaline. 
We see opportunities for research that explores how to reduce facilitators' workload and enhance the scalability of collective social support efforts through conversational agents capable of augmenting facilitator responsibilities. We draw on literature on human-AI collaboration \cite{grudin19,jiang2020}, which posits that humans and conversational agents have different strengths and limitations in online activities. Complementing the strengths of facilitators with those of conversational agents could enable expansion of the collective social support system. \nicki{For example, extending the asymmetrical design \cite{rahman2021}, conversational agents could effectively assist facilitators with information management tasks, while routine coordination tasks (e.g., translation and nuanced communications) are done by  facilitators}. One approach to facilitate this collaboration would be implementing the conversational agent on top of WhatsApp through their API. Conversational agents could propose automated responses based on teacher, caseworker, and expert requests. Facilitators would have the option to either choose the agent's response or craft their own.

\subsection{Hybrid Collective Social Support \& Technostress}\label{5-3}
Saharaline's collective social support intervention utilized smartphone-based WhatsApp. However, a major stressor in our findings (and in prior work \cite{varanasi2021Investigatinga}) is technostress. Here, we discuss the impact of deploying a support intervention using the same technology that was also a source of stress for teachers. In our intervention, teachers had the option of receiving support either through WhatsApp or in person. In our findings, several teachers chose to interact and receive support in person
, while also receiving assistive information (e.g., support materials) through WhatsApp. We postulate that an important rationale behind this strategy was to safeguard themselves from  the perceived stress associated with the excessive use of WhatsApp and other personal apps for work purposes. \citet{Ayyagari_2011} classifies this form of stress as techno-overload. Enabling teachers to have conversations about their problems in person is what made Saharaline truly hybrid as, at any given point, Saharaline was simultaneously functioning in offline spaces (between caseworkers and teachers) and online spaces (between caseworkers, experts, and facilitators).

Additionally, by intentionally choosing offline spaces for conversations, teachers could control where these conversations occurred. In section \ref{5-1-Challenges} teachers deliberately chose safe or private spaces. This choice gave them a sense of control and enabled them to keep work-related conversations, especially about their problems, out of their personal lives. This ability to control conversations could have been reduced had the caseworkers compelled teachers to have conversations on WhatsApp, leading to a form of technostress called techno-invasion \cite{Tarafdar2019}. The overall flexibility and agency for teachers to control how they received social support influenced their likelihood of implementing the provided solutions and returning for further support.

\section{Limitations, Future Work \& Conclusion}

This paper reports on preliminary experiences of implementing collective social support---a community-centric social support practice---aimed explicitly at workers in low-resource settings to improve their occupational well-being. We studied these experiences by designing and deploying Saharaline, a WhatsApp-based hybrid collective social support intervention that leveraged support organizations (educational non-profits) operating in low-resource Indian schools to assist teachers in addressing various stress-inducing challenges.

This study was exploratory and has the usual limitations posed by qualitative research. To understand the feasibility of the service and the experience of multiple stakeholders, we kept our sample size of teacher and support organization staff small. To augment our qualitative findings, we suggest that future research engage with a larger sample of teachers across a longer period (e.g., an academic year) to understand the relative duration in teachers' work lives where they find collective support helpful. Future studies could also focus on measuring teachers' perceived social support and its impact on different indicators of occupational well-being, such as technostress, work satisfaction, and burnout, which could further emphasize the importance of social support services for vulnerable working communities.  

While our findings with teachers indicate the effectiveness of collective social support in their work lives, it is important to understand the impact of such support on other work practices that fall under the larger umbrella of emotional labor. Work practices vary greatly between different professions and specific support practices that are effective for certain professions can be ineffective for others. Further research could generalize collective social support practices to improve occupational well-being outcomes across vulnerable professions. 

Lastly, we intentionally designed Saharaline to be simple in its implementation to understand the fundamental sociotechnical practices necessary to provide collective social support rooted in local community needs. Based on these preliminary findings, we plan to study the integration of a more robust sociotechnical system that leverages mature technological capabilities (e.g., use of conversational agents) to assist in coordination (intimating deadlines) as well as knowledge organization (organizing case files), and synthesizing activities (sharing summaries). Taken together, these future research directions could establish well-balanced sociotechnical collective social support practices that can increase the resilience of vulnerable working communities in low-resource settings.

\begin{acks}
This work was made possible by the generous assistance of various support organizations. We express special gratitude to Meghshala for their enduring support throughout our research. We extend our thank you to the teachers who participated in the study and shared their valuable experiences of using the intervention. Finally, we also thank the external reviewers for their  feedback on the paper.
\end{acks}
\bibliographystyle{ACM-Reference-Format}
\bibliography{mainbib}
\newpage
\appendix
\onecolumn 

\section{Distribution of Support Organizations} 
\begin{table} [htbp]
\begin{tabular}[t]{|p{.7in}|p{1.1in}|p{.8in}|p{.5in}|p{1.1in}|p{.75in}|}
\hline
         \textbf{School Name} & \textbf{State} & \textbf{Support Organization}  & \textbf{Focus}  & \textbf{Support Organization} & \textbf{Focus} \\
\hline
School-1 \& 2 &
Meghalaya (rural) &
Support Org 1 &
Teachers &
Support Org 2 &
Higher management

\\
\hline
  School-3 \& 4 &
Gujarat (semi-urban) &
Support Org 3 &
Students &
Support Org 2 &
Higher management 

\\
\hline
      School-5 \& 6 &
Karnataka (urban) &
Support Org 1 &
Teachers &
Support Org 4 &
Teachers 

\\
\hline

    \end{tabular}
    \caption{Table showing distribution of the support organizations that provided collective social support on Saharaline}
    \label{tab:sup-org}
\end{table}

\section{Scenario-based Design Activities} 
\begin{table} [htbp]

\begin{tabular}[t]{|p{.5in}|p{1.0in}|p{1.5in}|p{.7in}|p{1in}|}
\hline
         \textbf{Week} & \textbf{Activities} & \textbf{Probes}  & \textbf{Format}  & \textbf{Outcomes} \\
\hline
Week-1 & 
Literature and experience sharing  & 
Powerpoint presentation, group reading & 
Seminar &
Story scenarios \\
\hline
        Week-2 &
Needs assessment &
Story scenarios, experience narratives &
Co-creation workshop &
Problem scenarios \\
\hline
        Week-3 &
Brainstorming design principles and scenarios &
Expert walk through of problem scenarios, authors' prior research insights &
Co-creation workshop &
Design principles \\
\hline

Week-4 &
Intervention design &  
Powerpoint presentation, affinity diagramming &
Seminar &
Intervention architecture \\
\hline

    \end{tabular}
    \caption{Weekly breakdown of different activities and probes used in the design process.}
    \label{tab:design-activity}
\end{table}

\end{document}